\documentclass[10pt,journal,twoside,twocolumn,nofonttune]{IEEEtran}

\usepackage{tabularx}
\usepackage{siunitx}
\usepackage{graphicx}
\usepackage{amsmath}
\usepackage{mathrsfs}
\usepackage{amsbsy}
\usepackage{amssymb}
\usepackage{mathtools}
\usepackage{upgreek}
\usepackage[twoside]{rotating}
\usepackage{psfrag}
\usepackage[numbers]{natbib}
\usepackage{color,soul}

\definecolor{LightGray}{gray}{0.8}
\definecolor{Orange}{rgb}{1.0, 0.31, 0.0}
\definecolor{Green}{rgb}{0.3, 1.0, 0.3}
\definecolor{Blue}{rgb}{0.75,0.75,1}

\newcommand{\fig}[1]{Fig.~\ref{#1}}
\newcommand{\figs}[1]{Figs~\ref{#1}}
\newcommand{\tab}[1]{Table~\ref{#1}}

\newcommand{\bea}{\begin{eqnarray}}
\newcommand{\beal}[1]{\begin{eqnarray}\label{#1}}
\newcommand{\eea}{\end{eqnarray}}
\def\balg#1#2\ealg{\begin{align}\label{#1}#2\end{align}}
\def\balgnl#1\ealgnl{\begin{align*}#1\end{align*}}
\def\bma#1\ema{\begin{matrix}#1\end{matrix}}

\newcommand{\x}{{\mathbf r}}
\newcommand{\xt}{{\boldsymbol \rho}}

\renewcommand{\H}{{\mathbf H}}

\newcommand{\B}{{\mathbf B}}

\newcommand{\V}{{\mathbf V}}
\newcommand{\R}{{\mathbb R}}
\newcommand{\N}{{\mathbb N}}

\newcommand{\I}{{\mathbb I}}
\newcommand{\G}{{\mathbb G}}

\newcommand{\A}{{\mathbb A}}

\newcommand{\e}{\boldsymbol{\epsilon}}
\newcommand{\m}{\boldsymbol{\mu}}

\newcommand{\uphi}{{\mathbf e}_\varphi}

\newcommand{\uz}{{\mathbf e}_z}
\newcommand{\urho}{{\mathbf e}_\rho}

\begin{document}

\title{Oblique Multiple Scattering by Gyrotropic Cylinders}

\author{Grigorios P. Zouros, Konstantinos Delimaris, Carsten Rockstuhl, and Georgios D. Kolezas
\thanks{
C.R. acknowledges support by the Deutsche Forschungsgemeinschaft (DFG, German Research Foundation) under Germany's Excellence Strategy via the Excellence Cluster 3D Matter Made to Order (EXC-2082/1-390761711) and from the Carl Zeiss Foundation via the CZF-Focus@HEiKA Program.

Grigorios P. Zouros, Konstantinos Delimaris and Georgios D. Kolezas are with the School of Electrical and Computer Engineering, National Technical University of Athens, 15773 Athens, Greece (e-mail: zouros@ieee.org; k.delimaris@gmail.com; geokolezas@central.ntua.gr).

Carsten Rockstuhl is with the Institute of Theoretical Solid State Physics, Karlsruhe Institute of Technology, 76131 Karlsruhe, Germany; also with the Institute of Nanotechnology, Karlsruhe Institute of Technology, 76131 Karlsruhe, Germany (e-mail: carsten.rockstuhl@kit.edu).
}}

\markboth{}%
{ZOUROS \lowercase{{\itshape et al}}.: Oblique Multiple Scattering by Gyrotropic Cylinders}

\maketitle

\begin{abstract}
In this work, we develop a full-wave vectorial solution for the 2.5-dimensional (2.5-D), i.e., at oblique plane wave incidence, electromagnetic (EM) multiple scattering (MS) by a collection of gyrotropic cylinders. All cylinders are infinitely long and share a common $z$-axis. However, each cylinder can have a different cross-section with an arbitrary shape and different gyrotropic material properties, i.e., both gyroelectric and gyromagnetic anisotropies are considered. The solution to the problem combines the following three elements: (i) development of a superpotentials-based cylindrical vector wave function (CVWF) expansion to express the EM field in the gyrotropic region; (ii) utilization of the extended boundary condition method (EBCM) to account for non-circular cylinders; (iii) use of Graf's formulas, specifically adapted for the CVWFs, to apply the EBCM at each cylinder. The developed theory allows us to calculate various scattering characteristics, including the scattering and extinction cross-sections and the multipole decomposition, enabling the design and in-depth investigation of various contemporary engineering and physics applications. The method is exhaustively validated with analytical techniques and COMSOL Multiphysics. The computational performance is also discussed. Finally, we study a potential microwave application of the MS by ferrite configurations, and demonstrate broadband forward scattering by introducing oblique incidence and anisotropy. Our method may be used to analyze, design, and optimize contemporary microwave, optical, and photonic applications by beneficially tailoring the scattering properties via oblique incidence and anisotropy.
\end{abstract}


\IEEEpeerreviewmaketitle

\section{Introduction}

\IEEEPARstart{R}{ecent} advancements in the study of electromagnetic (EM) multiple scattering (MS) by cylindrical structures have showcased diverse approaches and applications, such as tunability of light transport in disordered systems \cite{Arruda2016-ft}, scattering invisibility and free-space field enhancement using all-dielectric nanoparticles \cite{Liu2017-ss}, effective medium descriptions for multilayer metamaterials \cite{Mavidis2023-lv}, shielding effectiveness using metamaterial cylindrical arrays \cite{kou_zou_tsi_24,Zouros2025-bc}, and oligomer-based highly directional switching \cite{Loulas2025-ut}.

Most past works on MS refer to isotropic cylinders under normal illumination. For instance, initial work considered the MS by two cylinders and was presented in \cite{Yousif1990-hg}. Four different techniques, including a boundary value method (BVM), an iterative method, a combined hybrid exact-method-of-moments (MoM), and a high-frequency approximation technique, have been utilized for a random set of circular perfect electric conducting (PEC) or dielectric cylinders in \cite{Elsherbeni1994-tx}. The MS by a finite number of arbitrary cross-section cylinders has been explored in \cite{Felbacq1994-zg} using the scattering matrix. Second-order corrections using Lax's quasicrystalline approximation have been applied in \cite{Linton2005-cy} for the MS by a random configuration of circular cylinders, while MS considering cylinders in two dielectric half-spaces has been studied in \cite{Pawliuk2013-pv} by a plane wave decomposition method. Dielectric plasma circular arrays have been examined in \cite{Wu2014-la} via a BVM, while a collocation multipole method has been developed in \cite{Lee2014-nn} for circular boundaries. More recently, the method of auxiliary sources (MAS) and the fast multipole method (FMM) have been combined for large arrays of circular cylinders \cite{Mastorakis2022-sf}, a T-matrix approach has been presented for arbitrarily shaped cross-sectional cylinders \cite{Rubio2022-tt}, as well as for beam synthesis \cite{Rimpilainen2023-ml}, while a shooting and bouncing rays numerical method has been developed for magnetodielectric circular cylinders \cite{Cho2023-fq}.

On the other hand, the works on MS by isotropic cylinders under oblique illumination are scarce and include the circular dimer case \cite{Yousif1988-tp, Yousif1988-tn} and arrays of circular cylinders \cite{Lee1990-oy, Henin2007-io}, all solved using scalar cylindrical eigenfunction expansions. In addition, multiple circular scatterers have been examined in \cite{Frezza2020-fq} via the Foldy-Lax MS equations. At the same time, a recursive aggregated centered T-matrix algorithm has been applied for a cluster of cylinders \cite{Degen2023-gl}.

While most of the above works focus on isotropic scatterers, studies on anisotropic ones are limited. The MS from an array of ferrite, i.e., gyromagnetic, circular cylinders under normal illumination has been examined in \cite{Kumar2015-yi} using scalar cylindrical expansions. Although gyrotropic cylinders permit TE/TM separation under normal excitation, the oblique illumination renders the problem more challenging because, inside the gyrotropic regions, the longitudinal components of the EM field do not satisfy homogeneous Helmholtz equations. Such a problem was studied in \cite{Okamoto1979-tn} for gyrotropic cylinders via the reciprocity theorem, with numerical results limited to circular ferrite cylinders in a specific finite periodic, i.e., nonrandom, configuration. Finally, isolated multilayered circular gyroelectric cylinders have been considered in \cite{Kharton2025-ep}, under oblique illumination, via an operator scattering theory.

The oblique illumination constitutes a 2.5-dimensional (2.5-D) scattering problem where the EM field is written as ${\mathbf F}(\x)={\mathbf F}(\xt)e^{i\beta z}$, ${\mathbf F}={\mathbf E},{\mathbf H}$, with $\x=(\rho,\varphi,z)$ the position vector in cylindrical coordinates, $\xt=(\rho,\varphi)$ the position vector in polar coordinates, and $\beta$ the propagation constant. When $\partial/\partial z\rightarrow i\beta\neq0$, the field components $E_z$ and $H_z$ in a gyrotropic medium are coupled and do not satisfy homogeneous Helmholtz equations. Analytical solutions for this case were constructed in the past using the theory of superpotentials, initially developed in \cite{Epstein1956-wd} for a gyromagnetic medium, and later in \cite{Przeziecki1979-sm} for both gyroelectric and gyromagnetic, i.e., gyrotropic, media, with applications on half-plane diffraction \cite{Hurd1985-ne} and gyrotropic resonators \cite{Movchan2012-jx}.

In this work, we develop a full-wave vectorial method for the oblique 2.5-D MS by a collection of non-circular parallel gyrotropic cylinders. The cylinders have an infinite length along a common $z$-axis, but each one may have a different cross-section and gyrotropic properties. The solution is based on the following novel points. (i) Newly developed cylindrical vector wave functions (CVWFs) based on the theory of superpotentials, and abbreviated hereafter as SUPER-CVWFs, that enable the vectorial expansion of the EM field in a gyrotropic medium when $\partial/\partial z\rightarrow i\beta\neq0$. (ii) To account for non-circular cross-sections, we utilize the extended boundary condition method (EBCM), formerly employed for the study of hybrid wave propagation in non-circular isotropic optical fibers \cite{Delimaris2025-ig}, and extend its solution by incorporating the SUPER-CVWF expansions in the kernels of the integral representations (IRs). A new numerical scheme is then developed that requires the contour integration along the boundary of each scatterer, where the integrand functions incorporate the gyrotropic properties of the latter. The application of the EBCM at each cylinder is achieved using Graf's formulas \cite{Felbacq1994-zg}, which we adapt specifically for the CVWFs to maintain the vectorial nature of the solution. (iii) An efficient full-wave solution for gyroelectric and gyromagnetic material properties, allowing for the analysis and design of microwave, optical, and photonic applications.

The method is exhaustively validated: first, with the analytical solution for two isotropic scatterers \cite{Yousif1988-tp}; second, with COMSOL Multiphysics for configurations consisting of multiple anisotropic scatterers. Both TE and TM oblique plane wave incidence is considered, while scatterers of various cross-sections---including circular, elliptical, rounded triangular---and various array setups---on axis, random distributions---are examined. The computational performance is also discussed.

\begin{figure}[!t]
	\centering
	\includegraphics[scale=0.95]{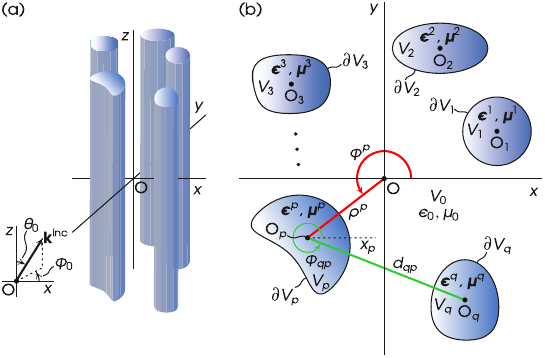}
	\caption{MS under oblique illumination. (a) 3-D view, (b) 2-D view in the $xy$-plane.}
	\label{geometry}
\end{figure}

The paper is organized as follows: Section~\ref{SOL1} discusses the setup of the MS problem, Section~\ref{SOL2} develops the SUPER-CVWFs, and Section~\ref{SOL3} develops the solution. Section~\ref{VAL} presents the validation and efficiency of the method, while Section~\ref{APP} discusses a microwave application involving broadband forward scattering by ferrite configurations. Finally, Section~\ref{CON} concludes the paper, while Appendices~\ref{APP_A} and \ref{APP_B} contain, respectively, Graf's formulas for the CVWFs and the necessary integrals for the construction of the system matrix.

\section{Problem Description}\label{SOL1}

The considered MS configuration is presented in \fig{geometry}, where a set of cylindrical scatterers is depicted. In \fig{geometry}(a), a three-dimensional (3-D) view is given, and the global coordinate system ${\rm O}xyz$ is defined. All cylinders have a common $z$-axis and are located in free space. The incident plane wave is described by the wavevector ${\mathbf k}^{\rm inc}$ and impinges obliquely at an angle $\theta_0$ with respect to the positive $z$-axis and at an angle $\varphi_0$ with respect to the positive $x$-axis. Two types of incidence are considered: TE and TM, where the vector of the incident magnetic and electric field, respectively, is parallel to the plane defined by ${\mathbf k}^{\rm inc}$ and the $z$-axis.

In \fig{geometry}(b), we define the global two-dimensional (2-D) coordinate system ${\rm O}xy$. Assuming a total of $P$ cylinders, a local coordinate system ${\rm O}_px_py_p$ is attached to each cylinder, where the location of the center ${\rm O}_p$, $p=1,2,\ldots,P$, is defined by the polar coordinates $(\rho^p,\varphi^p)$, $\rho^p\geqslant0$, $0\leqslant\varphi^p<2\pi$, with respect to ${\rm O}xy$. Each cylinder occupies a domain $V_p$ with boundary $\partial V_p$ given by the polar equation $\rho_p=\rho_p(\varphi_p)$, with $(\rho_p,\varphi_p)$, $\rho_p\geqslant0$, $0\leqslant\varphi_p<2\pi$, the polar coordinates with respect to ${\rm O}_px_py_p$. The relative distance between the centers ${\rm O}_q$ and ${\rm O}_p$ of two cylinders is denoted as $d_{qp}$, $q,p=1,2,\ldots,P$, $q\neq p$, while their relative angle $\varphi_{qp}$ is determined counterclockwise (CCW) from the positive ${\rm O}_px_p$ axis up to the line segment defined by $d_{qp}$, as depicted in \fig{geometry}(b). Finally, each cylinder is characterized by the gyrotropic permittivity and permeability tensors $\boldsymbol{\epsilon}^p$ and $\boldsymbol{\mu}^p$, given by
\balg{1}
\boldsymbol{\zeta}^p=\zeta_0\begin{bmatrix}
\zeta^p&-i\zeta_a^p&0\\
i\zeta_a^p&\zeta^p&0\\
0&0&\zeta_z^p
\end{bmatrix}\, ,\quad\boldsymbol{\zeta}^p=\boldsymbol{\epsilon}^p,\boldsymbol{\mu}^p\, ,
\ealg
where $\zeta^p=\epsilon^p,\mu^p$, $\zeta_a^p=\epsilon_a^p,\mu_a^p$, $\zeta_z^p=\epsilon_z^p,\mu_z^p$ are relative values and $\zeta_0=\epsilon_0,\mu_0$ the permittivity and permeability of free space, with the latter denoted as $V_0$ in \fig{geometry}(b). Throughout this paper, an $\exp(-i\omega t)$ time dependency is adopted.

\section{SUPER-CVWFs}\label{SOL2}

Following the analysis in \cite{Przeziecki1979-sm}, in a gyrotropic medium characterized by \eqref{1}, when $\partial/\partial z\neq0$, the components $E_z$ and $H_z$ satisfy, in stacked notation, the coupled equations
\balg{2}
\left(\nabla_t^2+\bma\epsilon_z/\epsilon\\\mu_z/\mu\ema\frac{\partial^2}{\partial z^2}+\bma k_e^2\\k_m^2\ema\right)\bma E_z(\x)\\H_z(\x)\ema=\mp\omega\bma\mu_z\\\epsilon_z\ema\tau_g\bma\partial H_z(\x)/\partial z\\\partial E_z(\x)/\partial z\ema,
\ealg
where $\nabla_t^2$ is the transverse Laplacian, $k_e=k_0(\epsilon_z\mu_\perp)^{1/2}$, and $k_m=k_0(\mu_z\epsilon_\perp)^{1/2}$, with $k_0=\omega(\mu_0\epsilon_0)^{1/2}$ the wavenumber of free space, $\mu_\perp=(\mu^2-\mu_a^2)/\mu$, $\epsilon_\perp=(\epsilon^2-\epsilon_a^2)/\epsilon$, and $\tau_g=\epsilon_a/\epsilon+\mu_a/\mu$. Introducing the generalized scalar Hertz potentials $u(\x)$ and $v(\x)$ by $E_z(\x)=-\epsilon/\epsilon_z\nabla_t^2u(\x)$, $H_z(\x)=-\mu/\mu_z\nabla_t^2v(\x)$, it can be shown \cite{Przeziecki1979-sm} that the EM field in a gyrotropic medium is expressed by
\balg{3}
\!\!\!\!\bma{\mathbf E}(\x)\\{\mathbf H}(\x)\ema\!=\!\bma\e^{-1}\\\m^{-1}\ema\nabla\!\times\!\bma\e\\\m\ema\nabla\!\times\!\bma u(\x)\\v(\x)\ema\uz \pm i\omega\bma\mu_0\mu/\epsilon\\\epsilon_0\epsilon/\mu\ema\bma\e^T\\\m^T\ema\nabla\!\times\!\bma v(\x)\\u(\x)\ema\uz\, ,
\ealg
where $T$ denotes transposition. Unlike the Hertz potentials for an isotropic medium, $u(\x)$ and $v(\x)$ do not satisfy the homogeneous Helmholtz equation. To overcome this obstacle, we introduce the so-called superpotentials $U$ and $V$ through
\balg{4}
\bma u(\x)\\v(\x)\ema&=\mp\omega\bma\mu\\\epsilon\ema\tau_g\bma\partial U(\x)/\partial z\\\partial V(\x)/\partial z\ema,\notag\\
\bma v(\x)\\u(\x)\ema&=\bma\epsilon/\epsilon_z\\\mu/\mu_z\ema\Big(\nabla_t^2+\bma\epsilon_z/\epsilon\\\mu_z/\mu\ema\frac{\partial^2}{\partial z^2}+\bma k_e^2\\k_m^2\ema\Big)\bma U(\x)\\V(\x)\ema\, .
\ealg
Replacing $\partial/\partial z\rightarrow i\beta$ and $\partial^2/\partial z^2\rightarrow -\beta^2$, further manipulations reveal that $U(\x)$ and $V(\x)$ satisfy the following homogeneous equation:
\balg{5}
\left(\nabla_t^2+\chi_1^2\right)\left(\nabla_t^2+\chi_2^2\right)\bma U(\x)\\V(\x)\ema=0\, ,
\ealg
where $\chi_1$ and $\chi_2$ are given by \cite{Hurd1985-ne}
\balg{6}
&\!\!\!\chi_{\substack{1\\2}}\!=\!\bigg\{\!\frac{1}{2} \Big\{k_e^2-\beta^2\frac{\epsilon_z}{\epsilon} +k_m^2-\beta^2\frac{\mu_z}{\mu}\notag\\
&\!\!\!\!\pm\! \Big[\!\Big(k_e^2-\beta^2\frac{\epsilon_z}{\epsilon}-k_m^2+\beta^2\frac{\mu_z}{\mu}\Big)^2\!\!+\!4\beta^2\tau_g^2k_0^2\epsilon_z\mu_z\!\Big]^{1/2}\Big\}\!\!\bigg\}^{1/2}\!\!\!\!.
\ealg
One can employ only one superpotential---either $U(\x)$ or $V(\x)$---to proceed with the solution. If $U(\x)$ is selected, the 1$^\mathrm{st}$ and the 3$^\mathrm{rd}$ equations of \eqref{4} are utilized. If $V(\x)$ is used instead, the 2$^\mathrm{nd}$ and the 4$^\mathrm{th}$ equations of \eqref{4} should be used. In what follows, we choose $U(\x)$. Since two wavenumbers $\chi_1$ and $\chi_2$ are involved in \eqref{5}, the general solution may be written as $U(\x)=U_1(\x)+U_2(\x)$ \cite{Epstein1956-wd}, where $(\nabla_t^2+\chi_j^2)U_j(\x)=0$, $j=1,2$.

Substituting $U(\x)=U_1(\x)+U_2(\x)$ into the 1$^\mathrm{st}$ and 3$^\mathrm{rd}$ equations of \eqref{4}, and the resulting $u(\x)$ and $v(\x)$ into \eqref{3}, we can express the EM fields as ${\mathbf E}(\x)={\mathbf E}_1(\x)+{\mathbf E}_2(\x)$, ${\mathbf H}(\x)={\mathbf H}_1(\x)+{\mathbf H}_2(\x)$, where ${\mathbf E}_j(\x)$, ${\mathbf H}_j(\x)$ depend on the respective $U_j(\x)$. In doing so, we finally obtain the following matrix representation for ${\mathbf E}_j(\x)$ and ${\mathbf H}_j(\x)$, $j=1,2$:
\balg{7}
\bma{\mathbf E}_j(\x)\\{\mathbf H}_j(\x)\ema=
\begin{bmatrix}
\bma S_j\\M_j\ema & \bma T_j\\N\ema & 0 \\[4mm]
-\bma T_j\\N\ema & \bma S_j\\M_j\ema & 0 \\[4mm]
0 & 0 & \bma W_j\\R_j\ema
\end{bmatrix}
{\mathbf U}_j(\x)\, ,
\ealg
where ${\mathbf U}_j(\x)=[\nabla_t,1]^TU_j(\x)$ and $\nabla_t=[\partial/\partial\rho,1/\rho~\partial/\partial\varphi]^T$. The matrix elements in \eqref{7} are given by
\balg{8}
S_j&=\omega\mu_0\mu\Big[\Big(\tau_g-\frac{\epsilon_a}{\epsilon}\Big)\beta^2+\epsilon_a\mu_\perp k_0^2-\frac{\epsilon_a}{\epsilon_z}\chi_j^2\Big]\, ,\notag\\
T_j&=i\omega\mu_0\mu\Big(-\beta^2+\epsilon\mu_\perp k_0^2-\frac{\epsilon}{\epsilon_z}\chi_j^2\Big)\, ,\notag\\
W_j&=-i\omega\mu_0\mu\frac{\epsilon}{\epsilon_z}\tau_g\beta\chi_j^2\, ,\notag\\
M_j&=i\beta\Big[-\beta^2+(\epsilon\mu_a\tau_g+\epsilon\mu_\perp )k_0^2-\frac{\epsilon}{\epsilon_z}\chi_j^2\Big]\, ,\notag\\
N&=-\epsilon\mu\tau_g\beta k_0^2\, ,\notag\\
R_j&=\frac{\mu}{\mu_z}\Big(-\beta^2+\epsilon\mu_\perp k_0^2-\frac{\epsilon}{\epsilon_z}\chi_j^2\Big)\chi_j^2\, .
\ealg
We note that $N$ does not depend on $\chi_j$.

We next develop the SUPER-CVWFs that allow for the vectorial expansion of the EM field in a gyrotropic medium when $\partial/\partial z\rightarrow i\beta\neq0$. We begin with ${\mathbf E}_j(\x)$. Since $U_j(\x)$ satisfies a homogeneous Helmholtz equation, we expand it in scalar cylindrical eigenfunctions, i.e.,
\balg{9}
U_j(\x)=e^{i\beta z}\sum_{m=-\infty}^\infty C_{jm}J_m(\chi_j\rho)e^{im\varphi}\, ,
\ealg
where $C_{jm}$ is an expansion coefficient and $J_m(x)$ the Bessel function. Substituting \eqref{9} into \eqref{7}, we obtain after some algebra
\balg{10}
{\mathbf E}_j(\x)&=e^{i\beta z}\!\!\!\!\!\sum_{m=-\infty}^\infty \!\!\!\!C_{jm}e^{im\varphi}\notag\\
&\quad\times\Big\{
\Big[S_j\chi_jJ'_m(\chi_j\rho)+\frac{imT_j}{\rho}J_m(\chi_j\rho)\Big]\urho\notag\\
&\quad+\!\!\Big[\!\!-\!T_j\chi_jJ'_m(\chi_j\rho)\!+\!\frac{imS_j}{\rho}J_m(\chi_j\rho)\Big]\uphi\notag\\
&\quad+\!W_jJ_m(\chi_j\rho)\uz\Big\}\, ,
\ealg
where $J'_m(x)$ is the derivative of $J_m(x)$ with respect to the argument. We next introduce the CVWFs ${\mathbf M}_m^{(1)}(\chi_j,\x)$, ${\mathbf N}_m^{(1)}(\chi_j,\x)$, and ${\mathbf L}_m^{(1)}(\chi_j,\x)$ \cite{Han2008-fj}, i.e.,
\balg{11}
{\mathbf M}_{m}^{(1)}(\chi_j,\x)&=e^{i\beta z}e^{im\varphi} \!\Big[i\frac{m}{\rho} J_{m}(\chi_j\rho)\urho\!-\!\chi_jJ'_m(\chi_j\rho)\uphi\Big]\, ,\notag\\
k_j{\mathbf N}_{m}^{(1)}(\chi_j,\x)&=e^{i\beta z}e^{im\varphi} \!\Big[i\beta\chi_j J'_m(\chi_j\rho)\urho\!-\!\frac{m\beta}{\rho}J_m(\chi_j\rho)\uphi\notag\\
&\quad\quad\quad\quad\quad+\!\chi_j^2J_m(\chi_j\rho)\uz\Big]\, ,\notag\\
{\mathbf L}^{(1)}_{m}(\chi_j,\x)&=e^{i\beta z}e^{im\varphi} \!\Big[\chi_j J'_m(\chi_j\rho)\urho\!+\!i\frac{m}{\rho}J_m(\chi_j\rho)\uphi\notag\\
&\quad\quad\quad\quad\quad+\!i\beta J_m(\chi_j\rho)\uz\Big]\, ,
\ealg
where the superscript $(1)$ denotes that the functions $J_m(x)$ and $J'_m(x)$ are used in \eqref{11}, while the wavenumber $k_j$ multiplying ${\mathbf N}^{(1)}_m(\chi_j,\x)$ is set as $k_j=(\chi_j^2+\beta^2)^{1/2}$, so that the properties $\nabla\times{\mathbf M}^{(1)}_m(\chi_j,\x)=k_j{\mathbf N}^{(1)}_m(\chi_j,\x)$ and $\nabla\times{\mathbf N}^{(1)}_m(\chi_j,\x)=k_j{\mathbf M}^{(1)}_m(\chi_j,\x)$ hold. Given the CVWFs of \eqref{11}, we seek an expansion of the form
\balg{12}
\!\!\!\!{\mathbf E}_j(\x)\!=\!\!\!\!\!\sum_{m=-\infty}^\infty \!\!\!\!\!C_{jm}\!\big[\tilde{{\mathbf M}}_m^{(1)}\!(\chi_j,\x)\!+\!\tilde{{\mathbf N}}_m^{(1)}\!(\chi_j,\x)\!+\!\tilde{{\mathbf L}}_m^{(1)}\!(\chi_j,\x)\big],
\ealg
where $\tilde{{\mathbf M}}_m^{(1)}(\chi_j,\x)=\pi_j{\mathbf M}_m^{(1)}(\chi_j,\x)$, $\tilde{{\mathbf N}}_m^{(1)}(\chi_j,\x)=\sigma_jk_j{\mathbf N}_m^{(1)}(\chi_j,\x)$, $\tilde{{\mathbf L}}_m^{(1)}(\chi_j,\x)=\tau_j{\mathbf M}_m^{(1)}(\chi_j,\x)$, with $\pi_j$, $\sigma_j$, $\tau_j$ coefficients to account for gyrotropy. Substituting \eqref{11} into \eqref{12} and equating components with \eqref{10}, we obtain
\balg{13}
\rho\text{- and }\varphi\text{-components: }&T_j=\pi_j,\,\, S_j=i\beta\sigma_j+\tau_j\, ,\notag\\
z\text{-component: }&W_j=\sigma_j\chi_j^2+i\tau_j\beta\, .
\ealg
It is possible to solve \eqref{13} uniquely for $\pi_j$, $\sigma_j$, $\tau_j$, yielding
\balg{14}
\tilde{{\mathbf M}}_m^{(1)}(\chi_j,\x)&=T_j{\mathbf M}_m^{(1)}(\chi_j,\x)\, ,\notag\\
\tilde{{\mathbf N}}_m^{(1)}(\chi_j,\x)&=\frac{W_j-i\beta S_j}{\chi_j^2+\beta^2}k_j{\mathbf N}_m^{(1)}(\chi_j,\x)\, ,\notag\\
\tilde{{\mathbf L}}_m^{(1)}(\chi_j,\x)&=\frac{\chi_j^2S_j-i\beta W_j}{\chi_j^2+\beta^2}{\mathbf L}_m^{(1)}(\chi_j,\x)\, .
\ealg
The $\tilde{{\mathbf M}}_m^{(1)}(\chi_j,\x)$, $\tilde{{\mathbf N}}_m^{(1)}(\chi_j,\x)$, and $\tilde{{\mathbf L}}_m^{(1)}(\chi_j,\x)$ vectors in \eqref{14} are the electric SUPER-CVWFs used to expand ${\mathbf E}_j(\x)$ via \eqref{12} in a gyrotropic medium when $\partial/\partial z\rightarrow i\beta\neq0$.

The above procedure is repeated for ${\mathbf H}_j(\x)$, that can be expanded in the form
\balg{15}
\!\!\!\!{\mathbf H}_j(\x)\!=\!\!\!\!\!\sum_{m=-\infty}^\infty \!\!\!\!\!C_{jm}\!\big[\hat{{\mathbf M}}_m^{(1)}\!(\chi_j,\x)\!+\!\hat{{\mathbf N}}_m^{(1)}\!(\chi_j,\x)\!+\!\hat{{\mathbf L}}_m^{(1)}\!(\chi_j,\x)\big],
\ealg
where $\hat{{\mathbf M}}_m^{(1)}(\chi_j,\x)$, $\hat{{\mathbf N}}_m^{(1)}(\chi_j,\x)$, and $\hat{{\mathbf L}}_m^{(1)}(\chi_j,\x)$ are the magnetic SUPER-CVWFs which are given by \eqref{14} by replacing $T_j\rightarrow N$, $W_j\rightarrow R_j$, and $S_j\rightarrow M_j$.

\section{Solution of the MS}\label{SOL3}

\subsection{IRs and the EBCM}

The formulation of the MS problem is based on 2.5-D IRs of the fields \cite{Delimaris2025-ig}; therefore, the factor $e^{i\beta z}$ is omitted, and the fields and the CVWFs depend only on the polar coordinates. In total, $P$ coupled IRs should be constructed due to the collective contribution of the multiple scatterers. The IR for the $p^\mathrm{th}$ cylinder, $p=1,2,\ldots,P$, is written in the respective local system ${\rm O}_px_py_p$ and it reads
\balg{16}
\!\!\!\!{\mathbf E}^{\rm inc}(\xt_p)\!\!+\!\!{\mathbf E}^{\rm sc}_p(\xt_p)\!+\!\!\sum_{\substack{q=1\\q\neq p}}^P \!{\mathbf E}^{\rm sc}_{q}(\xt_p)\!\!=\!\!\begin{cases}{\mathbf E}^{\rm tot}(\xt_p),\!\,\xt_p\!\in\! V_0\, ,\\0,\!\,\xt_p\!\in\!V_p\, ,\end{cases}
\ealg
with
\balg{17}
&{\mathbf E}^{\rm sc}_p(\xt_p)=\notag\\
&\frac{i}{k_0}Z_0(k_0^2\mathbb{I}+\nabla_p\nabla_p^T) \!\!\!\!\!\!\oint\limits_{\xt'_p \in \partial V_p}\!\!\!\!\!\!\G_0(k_c;\xt_p,\xt'_p)[{\mathbf e}'_{np} \times \textbf{H}^{\rm tot}(\xt'_p)]{\rm d}\xt'_p\notag\\
&+\nabla_p\times \!\!\!\!\!\!\oint\limits_{\xt'_p \in \partial V_p}\!\!\!\!\!\!\G_0(k_c;\xt_p,\xt'_p)[{\mathbf e}'_{np} \times \textbf{E}^{\rm tot}(\xt'_p)]{\rm d}\xt'_p\, .
\ealg
In \eqref{16} and \eqref{17}, $\xt_p$ is the position vector in the polar coordinates of ${\rm O}_px_py_p$, ${\mathbf E}^{\rm inc}(\xt_p)$ is the incident electric field, ${\mathbf E}^{\rm sc}_p(\xt_p)$ is the electric field scattered from the $p^\mathrm{th}$ cylinder, ${\mathbf E}^{\rm sc}_{q}(\xt_p)$ are the respective electric fields scattered from each of the remaining $P-1$ cylinders (with $q=1,2,\ldots,P$ and $q\neq p$), and ${\mathbf E}^{\rm tot}(\xt_p)$ and ${\mathbf H}^{\rm tot}(\xt_p)$ are the total electric and magnetic fields in $V_0$. All fields are expressed in ${\rm O}_px_py_p$. ${\mathbf E}^{\rm sc}_p(\xt_p)$ is given by its IR \eqref{17}, where $Z_0=(\mu_0/\epsilon_0)^{1/2}$, $\I$ is the unity dyadic, $\nabla_p$ is the del operator expressed in ${\rm O}_px_py_p$, $\G_0(k_c;\xt_p,\xt'_p)$ is the tensorial form of the free-space Green's function, as defined in detail in \cite{Delimaris2025-ig}, $k_c=(k_0^2-\beta^2)^{1/2}$, and ${\mathbf e}_{np}$ is the outwards normal unit vector on $\partial V_p$.

The application of EBCM requires the expansion of the involved EM fields in terms of CVWFs. Starting with the incident electric field, its Cartesian components are given by
\balg{17b}
E_x^{\rm inc}&=( -\cos\theta_0 \cos\varphi_0c_{\rm TM} -\sin\varphi_0c_{\rm TE} )\notag\\
&\quad\,\,\, \times e^{ik_0x\sin\theta_0\cos\varphi_0}e^{ik_0y\sin\theta_0\sin\varphi_0}e^{i\beta z}\, ,\notag\\
E_y^{\rm inc}&=( -\cos\theta_0 \sin\varphi_0c_{\rm TM} +\cos\varphi_0c_{\rm TE}) \notag\\
&\quad\,\,\, \times e^{ik_0x\sin\theta_0\cos\varphi_0}e^{ik_0y\sin\theta_0\sin\varphi_0}e^{i\beta z}\, ,\notag\\
E_z^{\rm inc}&=\sin\theta_0c_{\rm TM} \notag\\
&\quad\,\,\, \times e^{ik_0x\sin\theta_0\cos\varphi_0}e^{ik_0y\sin\theta_0\sin\varphi_0}e^{i\beta z}\, .
\ealg
In \eqref{17b}, when $c_{\rm TE}=1$ and $c_{\rm TM}=0$, TE incidence is considered ($E_z^{\rm inc}=0$), denoted by $E_{x,y,z}^{\rm inc}\equiv ^{\rm TE}\!\!\!E_{x,y,z}^{\rm inc}$. When $c_{\rm TE}=0$ and $c_{\rm TM}=1$, TM incidence is considered ($H_z^{\rm inc}=0$), denoted by $E_{x,y,z}^{\rm inc}\equiv ^{\rm TM}\!\!\!E_{x,y,z}^{\rm inc}$. Based on these expressions, the incident electric field is expanded as
\balg{18}
^{\substack{{\rm TE}\\{\rm TM}}}{\mathbf E}^{\rm inc}(\xt_p)=\sum_{m=-\infty}^\infty e^{i\delta_p}\frac{i^{\substack{m+1\\m}}}{k_c}\bma{\mathbf M}^{(1)}_m(k_c,\xt_p)\\{\mathbf N}^{(1)}_m(k_c,\xt_p)\ema e^{-im\varphi_0}\, ,
\ealg
where $\delta_p=k_0\rho^p(\sin\theta_0\cos\varphi_0\cos\varphi^p+\sin\theta_0\sin\varphi_0\sin\varphi^p)$ is the phase at the center of each cylinder. ${\mathbf M}^{(1)}_m(k_c,\xt_p)$ and ${\mathbf N}^{(1)}_m(k_c,\xt_p)$ are given by \eqref{11} with $k_j\equiv k_0$.

The scattered field ${\mathbf E}^{\rm sc}_p(\xt_p)$ is expanded as
\balg{19}
\!\!\!{\mathbf E}^{\rm sc}_p(\xt_p)\!=\!\!\!\!\sum_{m=-\infty}^\infty \!\!\!\big[A_m^p{\mathbf M}^{(3)}_m(k_c,\xt_p)+B_m^p{\mathbf N}^{(3)}_m(k_c,\xt_p)\big]\, ,
\ealg
where $A_m^p$ and $B_m^p$ are unknown expansion coefficients to be determined, while the superscript $(3)$ in the CVWFs implies that $J_m(x)$ and $J'_m(x)$ in \eqref{11} are replaced by $H_m(x)$ and $H'_m(x)$, respectively, where $H_m(x)$ is the Hankel function of the first kind---the superscript $(1)$ is omitted for simplicity---and $H'_m(x)$ is the derivative of $H_m(x)$ with respect to the argument. The expansion \eqref{19} also holds for each ${\mathbf E}^{\rm sc}_q(\xt_q)$ in the local system ${\rm O}_qx_qy_q$, with $\xt_q$ the position vector in the polar coordinates of ${\rm O}_qx_qy_q$. Then, ${\mathbf E}^{\rm sc}_{q}(\xt_p)$ is obtained if ${\mathbf E}^{\rm sc}_q(\xt_q)$ is translated from ${\rm O}_qx_qy_q$ to ${\rm O}_px_py_p$, by means of Graf's formulas \eqref{A1} for the CVWFs (see Appendix~\ref{APP_A}).

Finally, the fields inside the $p^\mathrm{th}$ gyrotropic cylinder are expanded in terms of SUPER-CVWFs, using \eqref{12} and \eqref{15}, i.e.,
\balg{19b}
\bma{\mathbf E}_p(\xt_p)\\{\mathbf H}_p(\xt_p)\ema=\sum_{m=-\infty}^\infty&\sum_{j=1}^2 \,C_{jm}^p\bigg[\bma\tilde{{\mathbf M}}_m^{(1)p}(\chi_j^p,\xt_p)\\\hat{{\mathbf M}}_m^{(1)p}(\chi_j^p,\xt_p)\ema\notag\\
&+\bma\tilde{{\mathbf N}}_m^{(1)p}(\chi_j^p,\xt_p)\\\hat{{\mathbf N}}_m^{(1)p}(\chi_j^p,\xt_p)\ema+\bma\tilde{{\mathbf L}}_m^{(1)p}(\chi^p_j,\xt_p)\\\hat{{\mathbf L}}_m^{(1)p}(\chi_j^p,\xt_p)\ema\bigg]\, .
\ealg
In \eqref{19b}, $C^p_{jm}$ are unknown expansion coefficients, while the superscript $p$ in the SUPER-CVWFs is applied to all quantities involving material properties. For example, $\tilde{{\mathbf N}}_m^{(1)p}(\chi_j^p,\xt_p)$ is given from \eqref{14} as $\tilde{{\mathbf N}}_m^{(1)p}(\chi_j^p,\xt_p)=(W_j^p-i\beta S_j^p)/[(\chi_j^p)^2+\beta^2]k_j^p{\mathbf N}_m^{(1)}(\chi_j^p,\xt_p)$. $\tilde{{\mathbf M}}_m^{(1)p}(\chi_j^p,\xt_p)$ and $\tilde{{\mathbf L}}_m^{(1)p}(\chi_j^p,\xt_p)$ are given similarly by \eqref{14}. 

To proceed, we first consider the lower branch of \eqref{16} for $\xt_p$ inside the inscribed circle of $\partial V_p$. Making use of the continuity of the tangential components of the electric and magnetic field on $\partial V_p$, ${\mathbf e}'_{np}\times{\mathbf E}^{\rm tot}(\xt'_p)$ and ${\mathbf e}'_{np}\times{\mathbf H}^{\rm tot}(\xt'_p)$ in \eqref{17} are replaced by ${\mathbf e}'_{np}\times{\mathbf E}_p(\xt'_p)$ and ${\mathbf e}'_{np}\times{\mathbf H}_p(\xt'_p)$, respectively, for $\xt'_p\in\partial V_p$. Then, we substitute into \eqref{16} the CVWF expansions of the involved fields and the tensorial expansion of $\G_0$ for $\rho_p<\rho'_p$ \cite{Delimaris2025-ig}, apply the differential operators $k_0^2\mathbb{I}+\nabla_p\nabla_p^T$ and $\nabla_p\times$, and employ the orthogonality properties of the CVWFs. Thus, we finally obtain, in stacked notation, two sets of linear non-homogeneous equations for each type of incidence, involving $A^p_m$, $B^p_m$, and $C^p_{jm}$:
\balg{20}
&\sum_{\mu=-\infty}^\infty\sum_{j=1}^2 C^p_{j\mu}\bma P^p_{j\mu m}\\Q^p_{j\mu m}\ema\notag\\
&+\frac{4k_c^2}{k_0}\sum_{\substack{q=1\\q\neq p}}^P\sum_{l=\infty}^\infty\bma A^q_l\\B^q_l\ema(-1)^{l+m}e^{i(l-m)\varphi_{qp}}H_{l-m}(k_cd_{qp})\notag\\
&=
\begin{cases}
\bma
-e^{i\delta_p}\frac{4k_c}{k_0}i^{m+1}e^{-im\varphi_0}\\
0
\ema,&\text{TE incidence}\, ,\\
\bma
0\\
-e^{i\delta_p}\frac{4k_c}{k_0}i^{m}e^{-im\varphi_0}
\ema,&\text{TM incidence}\, .
\end{cases}
\ealg
In \eqref{20}, $P^p_{j\mu m}=iI^p_{2,j\mu m}-Z_0I^p_{3,j\mu m}$, $Q^p_{j\mu m}=iI^p_{1,j\mu m}-Z_0I^p_{4,j\mu m}$, where $I^p_{s,j\mu m}$, $s=1,2,3,4$, are integrals given in Appendix~\ref{APP_B}.

Next, we consider the upper branch of \eqref{16} and repeat the steps outlined above, where now $\xt_p$ lies outside the circumscribed circle of $\partial V_p$ and the tensorial
expansion of
$\G_0$ for $\rho_p>\rho'_p$ is used. This way, another two sets
of linear equations are ultimately obtained, connecting $A^p_m$ and $B^p_m$ with $C^p_{jm}$, i.e.,
\balg{21}
\bma A_m^p\\B_m^p\ema=\frac{k_0}{4k_c^2}\sum_{\mu=-\infty}^\infty\sum_{j=1}^2 C_{j\mu m}^p\bma U_{j\mu m}^p\\V_{j\mu m}^p\ema\, ,
\ealg
where again stacked notation is used and $U^p_{j\mu m}=i\tilde{I}^p_{2,j\mu m}-Z_0\tilde{I}^p_{3,j\mu m}$, $V^p_{j\mu m}=i\tilde{I}^p_{1,j\mu m}-Z_0\tilde{I}^p_{4,j\mu m}$, and $\tilde{I}^p_{s,j\mu m}$, $s=1,2,3,4$, are integrals defined in Appendix~\ref{APP_B}.

Equations~\eqref{20} and \eqref{21} constitute four infinite non-homogeneous sets for each type of incidence and, upon truncation, can be solved for $A_m^p$, $B_m^p$, and $C_{jm}^p$, $j=1,2$. However, we can reduce the number of sets by a factor of two upon substituting \eqref{21} into \eqref{20}. Then, $C_{jm}^p$, $j=1,2$, are calculated via
\balg{22}
&\sum_{\mu=-\infty}^\infty\sum_{j=1}^2 C^p_{j\mu}\bma P^p_{j\mu m}\\Q^p_{j\mu m}\ema+\sum_{\substack{q=1\\q\neq p}}^P\sum_{\mu=-\infty}^\infty\sum_{j=1}^2 C_{j\mu}^p\notag\\
&\times\sum_{l=-\infty}^\infty\bma U_{j\mu l}^q\\V_{j\mu l}^q\ema(-1)^{l+m}e^{i(l-m)\varphi_{qp}}H_{l-m}(k_c d_{qp})\notag\\
&=
\begin{cases}
\bma
-e^{i\delta_p}\frac{4k_c}{k_0}i^{m+1}e^{-im\varphi_0}\\
0
\ema,&\text{TE incidence}\, ,\\
\bma
0\\
-e^{i\delta_p}\frac{4k_c}{k_0}i^{m}e^{-im\varphi_0}
\ema,&\text{TM incidence}\, .
\end{cases}
\ealg
If $M$ is the upper limit for the truncation of the $m$ and $\mu$ indices in \eqref{22}, i.e., $m,\mu=-M,-(M-1),\ldots,-1,0,1,\ldots,M-1,M$, then the size of the system matrix for the calculation of $C_{jm}^p$ is $2P(2M+1)\times2P(2M+1)$. Once $C_{jm}^p$ are computed, $A_m^p$ and $B_m^p$ are evaluated from \eqref{21}.

\subsection{Cross-sections}

The scattering and extinction cross-sections $Q_{\rm sca}$ and $Q_{\rm ext}$ are important quantities for the analysis and design of various scattering configurations. To calculate $Q_{\rm sca}$ and $Q_{\rm ext}$ for a collection of scatterers, the locally defined ${\mathbf E}^{\rm sc}_p(\xt_p)$ in \eqref{19} is first translated to the global system ${\rm O}xy$. Applying \eqref{A2} and interchanging indices $m$ and $l$, we get
\balg{23}
\!\!\!{\mathbf E}^{\rm sc}_p(\xt)&=\sum_{m=-\infty}^\infty \!\!\!\big[\hat{A}_m^p{\mathbf M}^{(3)}_m(k_c,\xt)+\hat{B}_m^p{\mathbf N}^{(3)}_m(k_c,\xt)\big]\, ,\notag\\
\!\!\!\bma \hat{A}_m^p\\\hat{B}_m^p\ema&=(-1)^m\sum_{l=-\infty}^\infty \bma A_l^p\\B_l^p\ema (-1)^l e^{i(l-m)\varphi^p}J_{l-m}(k_c\rho^p)\, ,
\ealg
where $\xt$ is the position vector in polar coordinates in the global ${\rm O}xy$. Then, the total scattered field ${\mathbf E}^{\rm sc}(\xt)$ of the collection, with respect to ${\rm O}xy$, is given by
\balg{24}
{\mathbf E}^{\rm sc}(\xt)&=\sum_{m=-\infty}^\infty \!\!\!\big[A_m{\mathbf M}^{(3)}_m(k_c,\xt)+B_m{\mathbf N}^{(3)}_m(k_c,\xt)\big]\, ,\notag\\
\bma A_m\\B_m\ema&=\sum_{p=1}^P\bma \hat{A}_m^p\\\hat{B}_m^p\ema\, ,
\ealg
and $Q_{\rm sca}$ and $Q_{\rm ext}$ are calculated via \cite{bohren}
\balg{25}
Q_{\rm sca}&=\frac{4}{k_0}\sum_{m=-\infty}^\infty\big(|\tilde{A}_m|^2+|\tilde{B}_m|^2\big)\, ,\notag\\
Q_{\rm ext}&=\begin{cases}-\frac{4}{k_0}{\rm Re}\sum_{m=-\infty}^\infty\tilde{A}_m,&\text{TE incidence}\, ,\\-\frac{4}{k_0}{\rm Re}\sum_{m=-\infty}^\infty\tilde{B}_m,&\text{TM incidence}\, ,\end{cases}
\ealg
where ${\rm Re}$ denotes the real part. In \eqref{25} we have introduced the dimensionless coefficients $\tilde{A}_m$ and $\tilde{B}_m$, defined by $\tilde{G}_m=G_mk_c(-i)^{m+1}e^{im\varphi_0}$ for TE incidence and $\tilde{G}_m=G_mk_c(-i)^{m}e^{im\varphi_0}$ for TM incidence, with $\tilde{G}_m=\tilde{A}_m,\tilde{B}_m$ and $G_m=A_m,B_m$. 

\begin{figure}[!t]
\centering
\includegraphics[scale=0.95]{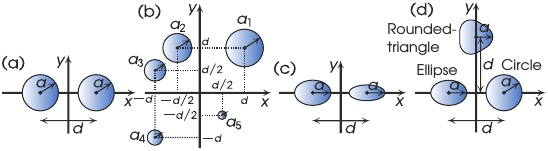}
\caption{
(a) Circular dimer, (b) randomly placed circular scatterers, (c) elliptical non-symmetric dimer, and (d) randomly placed scatterers of different shapes.
}
\label{cases}
\end{figure}

Finally, the scattering width (SW) $\sigma(\varphi)$, with $0\leqslant\varphi<2\pi$ the observation angle with respect to ${\rm O}xy$, for a collection of scatterers, is computed by
\balg{26}
\sigma(\varphi)=\frac{2\pi}{k_c}[|f_\varphi(\varphi)|^2+|f_z(\varphi)|^2]\, ,
\ealg
where
\balg{27}
\bma f_\varphi(\varphi)\\ f_z(\varphi)\ema=k_c\sqrt{\frac{-2i}{\pi}}\sum_{m=-\infty}^\infty (-i)^m e^{im\varphi}\bma B_m\\A_m\ema\, .
\ealg

\section{Validation and Performance}\label{VAL}

We demonstrate the validity and efficiency of our method for various MS configurations by performing comparisons with the analytical solution \cite{Yousif1988-tp} for isotropic scatterers, and with COMSOL for anisotropic ones. We consider both TM and TE oblique plane wave incidence, set $\theta_0=45^\circ$ and $\varphi_0=30^\circ$, and compute the normalized spectra $Q_{\rm sca}/\lambda_0$ and $Q_{\rm ext}/\lambda_0$ vs $k_0a$ for a variety of setups depicted in \fig{cases}, where $\lambda_0=2\pi/k_0$ is the free-space wavelength, and $a$ a geometrical parameter depicted in \fig{cases}, e.g., $a$ is the radius of the cylinders in \fig{cases}(a). As far as anisotropy is concerned, both gyroelectric and gyromagnetic setups are examined.

\begin{figure}[!t]
\centering
\includegraphics[scale=0.95]{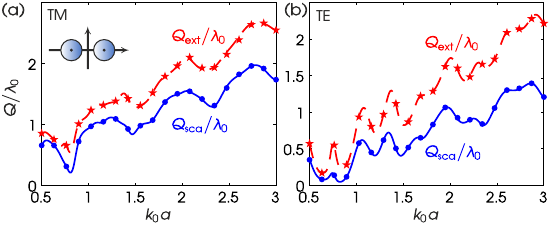}
\caption{
$Q_{\rm sca}/\lambda_0$ and $Q_{\rm ext}/\lambda_0$ vs $k_0a$ for the circular dimer of \fig{cases}(a). (a) TM incidence, (b) TE incidence. Values of parameters: $\epsilon^p=25+i2$, $\epsilon^p_a=0$, $\epsilon^p_z=25+i2$, $\mu^p=1$, $\mu^p_a=0$, $\mu^p_z=1$ ($p=1,2$), $\theta_0=45^\circ$, $\varphi_0=30^\circ$, and $d/a=3$. Blue solid curve/dots: $Q_{\rm sca}/\lambda_0$; red dashed curve/stars: $Q_{\rm ext}/\lambda_0$; curves: this work; symbols: \cite{Yousif1988-tp}.
}
\label{two-iso}
\end{figure}

At first, in \fig{two-iso}, we consider a circular dimer composed of two isotropic cylinders---see \fig{cases}(a). The values of parameters are gathered in the Figure's caption. This case can also be solved analytically \cite{Yousif1988-tp}. In \figs{two-iso}(a) and (b), we plot the normalized values of $Q_{\rm sca}/\lambda_0$ and $Q_{\rm ext}/\lambda_0$ vs $k_0a$, for TM and TE incidence, respectively. As is evident, the spectra obtained by the present method and \cite{Yousif1988-tp} are in absolute agreement.

\begin{table}[!b]
\caption{Computational performance.}
\label{tab1}
\centering
\vspace{-4mm}
\rule{\linewidth}{1pt}
\begin{tabular*}{\linewidth}{@{\extracolsep{\fill}}lS[table-format=3.0]S[table-format=4.0]S[table-format=1.3]S[table-format=1.2]}
& \multicolumn{2}{c}{CPU time~(s)} & \multicolumn{2}{c}{RAM~(GB)} \\
\cline{2-3}\cline{4-5} 
{Example} & {This work} & {COMSOL} & {This work} & {COMSOL} \\
\cline{1-5}
\fig{two-c}   & 19  & 908  & 0.064 & 2.37 \\
\fig{five-c}  & 98  & 1008 & 0.065 & 2.17 \\
\fig{two-ed}  & 242 & 822  & 0.064 & 2.25 \\
\fig{three-m} & 449 & 3299 & 0.064 & 4.2  \\

\end{tabular*}
\rule{\linewidth}{1pt}
\end{table}

\begin{figure}[!t]
\centering
\includegraphics[scale=0.95]{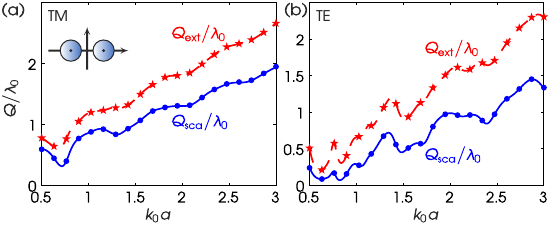}
\caption{
$Q_{\rm sca}/\lambda_0$ and $Q_{\rm ext}/\lambda_0$ vs $k_0a$ for the circular dimer of \fig{cases}(a). (a) TM incidence, (b) TE incidence. Values of parameters: $\epsilon^p=25+i2$, $\epsilon^p_a=4$, $\epsilon^p_z=30+i5$, $\mu^p=1$, $\mu^p_a=0$, $\mu^p_z=1$ ($p=1,2$), $\theta_0=45^\circ$, $\varphi_0=30^\circ$, and $d/a=3$. Blue solid curve/dots: $Q_{\rm sca}/\lambda_0$; red dashed curve/stars: $Q_{\rm ext}/\lambda_0$; curves: this work; symbols: COMSOL.
}
\label{two-c}
\end{figure}

In the remaining work, we focus entirely on anisotropic configurations and compare our method with COMSOL. For each example studied, we provide in \tab{tab1} the computational performance of our technique and compare it against COMSOL's. We begin with gyroelectric setups. \fig{two-c} examines a circular dimer---see \fig{cases}(a)---composed of two gyroelectric cylinders. The agreement between our method and COMSOL is evident. Each curve in \fig{two-c} is plotted using $251$ different values of $k_0a$; setting the truncation limit to $M=6$, it is seen from \tab{tab1} that our method requires $19$~s of CPU time vs the $908$~s of COMSOL, with the latter initialized using an extremely fine mesh. In addition, the $0.064$~GB memory consumption is much lower than the $2.37$~GB that COMSOL's finite-element solver requires. These data hold for either TM or TE incidence.

\begin{figure}[!t]
\centering
\includegraphics[scale=0.95]{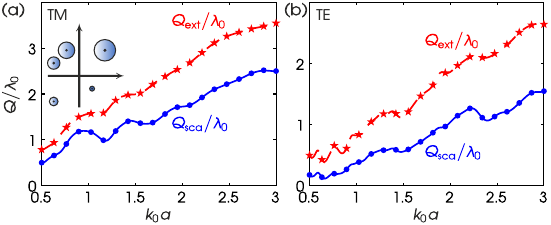}
\caption{
$Q_{\rm sca}/\lambda_0$ and $Q_{\rm ext}/\lambda_0$ vs $k_0a$ for the randomly placed circular scatterers of \fig{cases}(b). (a) TM incidence, (b) TE incidence. Values of parameters: same as in \fig{two-c} but $d/a=2.4$, and also $a_2/a_1=0.8$, $a_3/a_1=0.6$, $a_4/a_1=0.4$, $a_5/a_1=0.2$, with $a_1\equiv a$. Legends: same as in \fig{two-c}.
}
\label{five-c}
\end{figure}

In \fig{five-c}, we consider five randomly placed circular gyroelectric cylinders of different radius---see \fig{cases}(b). As is evident, the agreement between our method and COMSOL is excellent for the entire range of $k_0a$ used in the example. However, our method significantly outperforms COMSOL by a speed-up of $1,008~{\rm s}/98~{\rm s}\approx10.3$ times, thus rendering our technique appropriate for efficient calculations.

\begin{figure}[!t]
\centering
\includegraphics[scale=0.95]{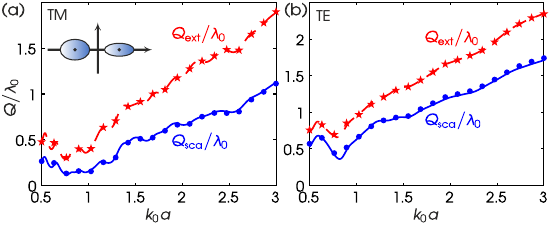}
\caption{
$Q_{\rm sca}/\lambda_0$ and $Q_{\rm ext}/\lambda_0$ vs $k_0a$ for the elliptical non-symmetric dimer of \fig{cases}(c). (a) TM incidence, (b) TE incidence. Values of parameters: $\epsilon^p=1$, $\epsilon^p_a=0$, $\epsilon^p_z=1$, $\mu^p=25+i2$, $\mu^p_a=4$, $\mu^p_z=30+i5$ ($p=1,2$), $\theta_0=45^\circ$, $\varphi_0=30^\circ$, $d/a=3$, ${\rm AR}_\text{left}=0.8$, and ${\rm AR}_\text{right}=0.6$. Blue solid curve/dots: $Q_{\rm sca}/\lambda_0$; red dashed curve/stars: $Q_{\rm ext}/\lambda_0$; curves: this work; symbols: COMSOL.
}
\label{two-ed}
\end{figure}

Gyromagnetic configurations are studied in the two remaining examples. In particular, \fig{two-ed} considers an elliptical non-symmetric dimer---see \fig{cases}(c). The left ellipse has an aspect ratio (AR), denoted as ${\rm AR}_{\rm left}$, given by ${\rm AR}_{\rm left}=(\text{minor semi-axis})/(\text{major semi-axis})=0.8$, where $a$ is the major semi-axis of the left ellipse. The right ellipse has the same major semi-axis but ${\rm AR}_{\rm right}=0.6$. Both scatterers are gyromagnetic, and the values of parameters are given in the Figure's caption. Our method and COMSOL follow each other for the entire $k_0a$ range. Nevertheless, the great merit of our technique is its efficiency, as it requires $242$~s to compute all $251$ $k_0a$ points of the spectrum, while COMSOL needs $822$~s.

\begin{figure}[!t]
\centering
\includegraphics[scale=0.95]{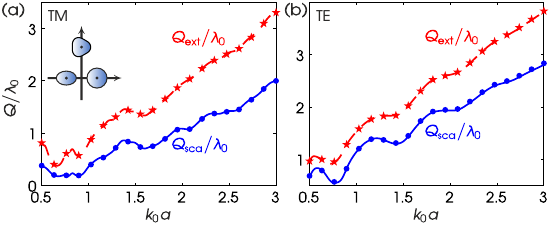}
\caption{
$Q_{\rm sca}/\lambda_0$ and $Q_{\rm ext}/\lambda_0$ vs $k_0a$ for the randomly placed scatterers of \fig{cases}(d). (a) TM incidence, (b) TE incidence. Values of parameters: same as in \fig{two-ed} with ${\rm AR}=0.8$. Legends: same as in \fig{two-ed}.
}
\label{three-m}
\end{figure}

Finally, in \fig{three-m}, we examine three randomly placed gyromagnetic scatterers of different shapes ---see \fig{cases}(d). The collection consists of a circular, an elliptical, and a rounded-triangle cylinder. The circular cylinder ($p=1$) has radius $a$, the elliptical cylinder ($p=2$) has a major semi-axis $a$ and ${\rm AR}=0.8$, while the rounded-triangle cylinder ($p=3$) is described by the polar equation $\rho_3(\varphi_3)=a[h^2+2h\cos(3\varphi_3)+1]^{1/2}/(h+1)$, $0\leqslant\varphi_3<2\pi$, with $h=0.1$. The agreement between our method and COMSOL is evident, thus establishing the correctness of our implementation. This specific example also showcases the efficiency of our algorithm since it requires $449$~s of CPU time to complete the calculations. In contrast, COMSOL needs $3,299$~s, a $7.3$ times speed-up. Moreover, our implementation requires a small amount of $0.064$~GB RAM while COMSOL requires $4.2$~GB, thus further revealing the efficiency of the proposed method.

\section{Broadband Huygens Sources by YIG Arrays}\label{APP}

In general, the design of contemporary scattering applications is commonly carried out under normal wave incidence, and the analysis is based on the multipole decomposition of the $Q_{\rm sca}$ spectrum. Herein, we employ our method to: (i) reveal how oblique incidence significantly affects the $Q_{\rm sca}$ spectrum and the multipole decomposition, giving rise to interesting phenomena, such as broadband Huygens sources; (ii) unveil how array configurations affect the $\sigma$ response; (iii) examine the role of anisotropy for the optimization of the system under study.

To this end, we study a microwave application involving oblique scattering by G-113 yttrium garnet (YIG) ferrite array configurations. In the presence of an external magnetic flux density bias ${\mathbf B}_0=B_0\uz$, YIG exhibits a gyromagnetic response \cite{pozar_4ed}: the $\m^p$ relative elements are $\mu^p=1+\omega_0 \omega_m/(\omega_0^2-\omega^2)$, $\mu_{a}^p=\omega\omega_m/(\omega_0^2-\omega^2)$, and $\mu_{z}^p=1$. In these relations, $\omega_m=\mu_0\gamma M_s$, $\gamma=1.759\times 10^{11} \mathrm{C/kg}$ is the gyromagnetic ratio, $4\pi M_s=1780$~G \cite{pozar_4ed}, $\omega_0=\mu_0\gamma H_0-i \mu_0 \gamma \Delta H/2$ is the complex Larmor circular frequency, $H_0=B_0/\mu_0$ is the external magnetic field intensity, and $\Delta H=45~\mathrm {Oe}$ \cite{pozar_4ed}. The $\e^p$ is isotropic with relative values $\epsilon^p=\epsilon_{z}^p=15(1+i0.0002)$ and $\epsilon_{a}^p=0$ \cite{pozar_4ed}.

\begin{figure}[!t]
\centering
\includegraphics[scale=0.95]{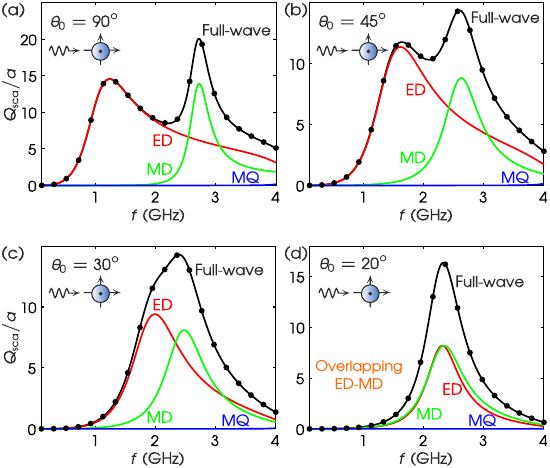}
\caption{
$Q_{\rm sca}/a$ vs $f$ for a single YIG cylinder under TM incidence. (a) $\theta_0=90^\circ$, (b) $\theta_0=45^\circ$, (c) $\theta_0=30^\circ$, (d) $\theta_0=20^\circ$. Values of parameters: $a=1$~cm, $\varphi_0=0^\circ$, $B_0=1$~T. Black curve: full-wave/this work; red curve: ED/this work; green: MD/this work; blue: MQ/this work; black dots: full-wave/COMSOL.}
\label{app1}
\end{figure}

In \fig{app1}, we consider a single YIG cylinder of radius $a=1$~cm, magnetized at $B_0=1$~T, and study how oblique TM incidence affects the $Q_{\rm sca}$ spectrum. The values of the parameters are given in the Figure's caption. \fig{app1}(a) depicts the normalized $Q_{\rm sca}/a$ response, along with its multipole decomposition, for $\theta_0=90^\circ$, i.e., normal incidence. The full-wave solution is obtained by \eqref{25}, keeping all the $m$-summation terms necessary for convergence. For TM incidence, the computation of $Q_{\rm sca}$ solely by the $m=0$ term gives the electric dipolar (ED) response, the $m=\pm1$ terms give the magnetic dipolar (MD) response, while the $m=\pm2$ terms yield the magnetic quadrupolar (MQ) response \cite{Liu2017-ss}, \cite{Loulas2025-ut}. We note that TE incidence is not considered in the context of this study since it does not yield the interesting phenomena that the TM case does, as discussed below, and also because at normal incidence the anisotropy does not have an impact on TE polarization \cite{kat_zou_rou_21}. The results of the full-wave solution plotted in \fig{app1} are in absolute agreement with COMSOL. In the frequency window $0.1$--$4$~GHz depicted in \fig{app1}, the higher-order MQ response is negligible. The ED and MD resonances, located at $1.23$~GHz and $2.72$~GHz, respectively, have a distinct separation. Next, in \figs{app1}(b)--(d), we introduce oblique incidence by setting in turn $\theta_0=45^\circ,30^\circ,20^\circ$. It is evident that the ED and MD resonances start to shift closer to each other and, surprisingly, for $\theta_0=20^\circ$ they overlap, not for a single but for a band of $f$, starting from $1$~GHz up to roughly $4$~GHz. The resonant peak in \fig{app1}(d) is located at $f_0=2.32$~GHz. The main conclusion of \fig{app1} is that, using oblique incidence excitation, one can beneficially tailor the spectrum by matching the ED and MD responses, in order to achieve interesting scattering phenomena, as discussed below. 

\begin{figure*}[!t]
\centering
\includegraphics[scale=0.95]{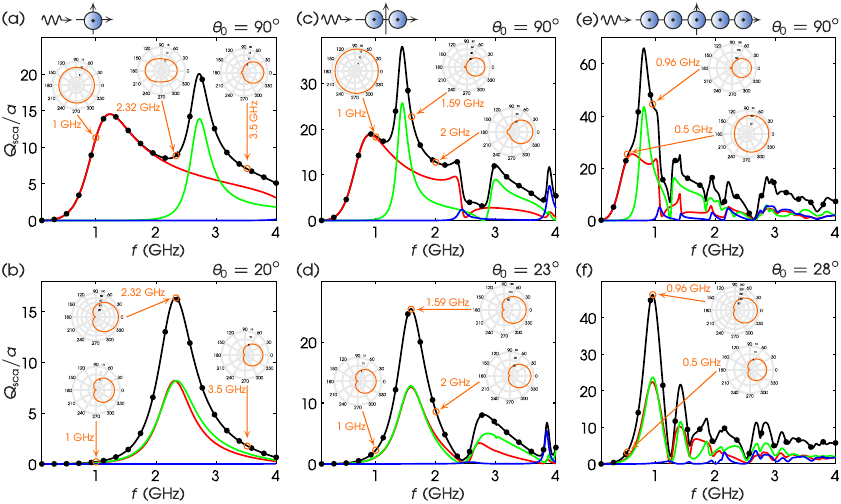}
\caption{
$Q_{\rm sca}/a$ vs $f$ for different YIG configurations under TM incidence. All cylinders have $a=1$~cm while $\varphi_0=0^\circ$.
(a)--(b) Single cylinder with $B_0=1$~T; (a) $\theta_0=90^\circ$; (b) $\theta_0=20^\circ$.
(c)--(d) Dimer with $d=2.4$~cm and $B_0=0.4$~T; (c) $\theta_0=90^\circ$; (d) $\theta_0=23^\circ$.
(e)--(f) Array of five cylinders with $d=2.4$~cm and $B_0=0.1$~T; (e) $\theta_0=90^\circ$; (f) $\theta_0=28^\circ$.
Black curve: full-wave/this work; red curve: ED/this work; green: MD/this work; blue: MQ/this work; black dots: full-wave/COMSOL.
Insets: $\sigma/a$ at the depicted $f$.
}
\label{app2}
\end{figure*}

Spectrum tailoring also holds true for array configurations of YIG. In \fig{app2}, we demonstrate how the MS mechanism allows for such a tailoring and how it affects the $\sigma$ response. To this end, \fig{app2} presents three studies. Study No~1/\figs{app2}(a) and (b): the single YIG cylinder examined in \fig{app1}; study No~2/\figs{app2}(c) and (d): a dimer of YIG cylinders; study No~3/\figs{app2}(e) and (f): an array of five YIG cylinders. The top row of \fig{app2} depicts the $Q_{\rm sca}/a$ at normal incidence for all setups, while the bottom row depicts the $Q_{\rm sca}/a$ at oblique incidence; we note that for each setup (single cylinder, dimer, array), the ED-MD overlap occurs for a different $\theta_0$ angle. It is evident that, under normal incidence, the locations of the ED and MD resonances are distinctly separated for all the examined setups. Indeed, the ED/MD resonances occur at $1.23$~GHz/$2.72$~GHz, $0.91$~GHz/$1.44$~GHz, and $0.62$~GHz/$0.81$~GHz for the single cylinder, the dimer, and the array setup, respectively. By applying oblique incidence, we achieve ED-MD overlapping, where the respective peaks of the resulting resonances are located at $f_0=2.32$~GHz, $f_0=1.59$~GHz, and $f_0=0.96$~GHz. What is more, this overlapping is observed along a wide range of frequencies: from \figs{app2}(b), (d), and (f), the respective ranges are $1$--$4$~GHz, $0.8$--$2.3$~GHz, and $0.3$--$1.25$~GHz.

The main impact of the ED-MD overlapping, and consequently of the oblique excitation, is the forward scattering achieved, while maintaining backward scattering at a minimum. To show this, we select three indicative frequencies inside the aforementioned frequency ranges where the ED-MD overlapping takes place. These frequencies are depicted in the insets of \fig{app2}. Then, for these specific $f$, we calculate the normalized $\sigma/a$, which are also plotted in the insets of \fig{app2}. As observed, the $\sigma/a$ for the $\theta_0=90^\circ$ cases do not exhibit interesting scattering characteristics, while the patterns are different for each considered $f$: e.g., in \fig{app2}(c), the $\sigma/a$ at $1$~GHz has a completely different pattern as compared to the $\sigma/a$ at $1.59$~GHz or at $2$~GHz. However, the respective $\sigma/a$ in \fig{app2}(d) shows that, for the same $f$, we achieve forward scattering with reduced backward scattering. In fact, this behavior is realized for all $f$ within $0.8$--$2.3$~GHz, as well as for all $f$ within $1$--$4$~GHz (single cylinder setup) and within $0.3$--$1.25$~GHz (array setup), thus showcasing that the system, for a suitable angle $\theta_0$, behaves like a Huygens scatterer for a broadband of $f$, where ED and MD contribute equally. In addition to the above observations, the MS mechanism operates collectively to enhance the maximum value of $\sigma/a$ at the forward scattering direction. Indeed, observing these values at $f_0=2.32$~GHz, $f_0=1.59$~GHz, and $f_0=0.96$~GHz in \figs{app2}(b), (d), and (f), we get the respective values of $\sigma(\varphi=0^\circ)/a=90.71$, $\sigma(\varphi=0^\circ)/a=139.38$, and $\sigma(\varphi=0^\circ)/a=237.26$.

\begin{figure}[!t]
\centering
\includegraphics[scale=0.95]{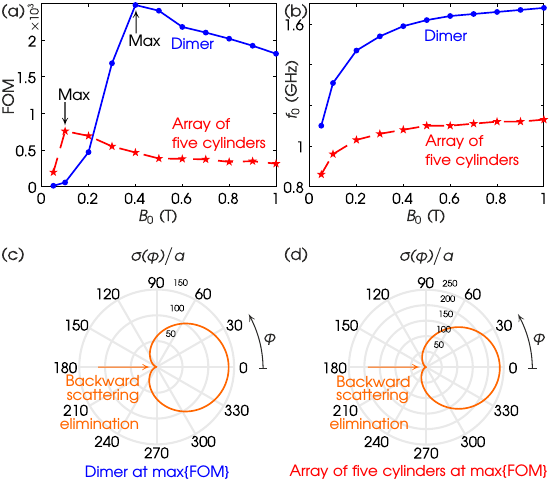}
\caption{
(a) FOM vs $B_0$. All YIG cylinders have $a=1$~cm, $d=2.4$~cm while $\varphi_0=0^\circ$. Two cylinders: $\theta_0=23^\circ$; five cylinders: $\theta_0=28^\circ$.
(b) $f_0$ vs $B_0$ for the same setups as in (a).
(c) $\sigma(\varphi)/a$ for the two cylinders of (a) at $\max\{\text{FOM}\}$ where $B_0=0.4$~T and $f_0=1.59$~GHz.
(d) $\sigma(\varphi)/a$ for the five cylinders of (a) at $\max\{\text{FOM}\}$ where $B_0=0.1$~T and $f_0=0.96$~GHz.
}
\label{app3}
\end{figure}

Finally, in \fig{app3}, we examine the role of anisotropy in optimizing the scattering response. For this purpose, we define the figure-of-merit (FOM) as ${\rm FOM}=\sigma(\varphi=0^\circ)/\sigma(\varphi=180^\circ)$, i.e., the front-to-back ratio with $\sigma(\varphi=0^\circ)$ the SW corresponding to a scattering into the strictly forward direction and $\sigma(\varphi=180^\circ)$ the SW corresponding to a scattering into the strictly backward scattering. Then we vary the gyromagnetic anisotropy of the YIG ferrite by changing the external bias $B_0$. In \fig{app3}(a), the optimal state ${\rm FOM}=2484$ is observed at $B_0=0.4$~T for the dimer where the FOM is maximized. The respective value for the array of five cylinders is ${\rm FOM}=759$ at $B_0=0.1$~T. Next, in \fig{app3}(b), we study how $f_0$, i.e., the frequency of the resonant peak when the ED and MD resonances overlap, varies with the change of $B_0$. We notice that the increase of the external bias causes $f_0$ to shift to larger values. In particular, $f_0=1.59$~GHz for the dimer at $\max\{{\rm FOM}\}$, while the respective value for the array is $f_0=0.96$~GHz. Therefore, $B_0$ may be used as an external switch allowing us to select $f_0$ to achieve forward scattering with maximized forward scattering behavior at different frequencies. Finally, in \figs{app3}(c) and (d), we plot the $\sigma/a$ for the two cases of \fig{app3}(a) where the FOM is maximized. Forward scattering is observed for both the dimer and the array setup, while the backward scattering is eliminated.

\section{Conclusion}\label{CON}

In this work, we developed a full-wave solution for the problem of EM MS of obliquely incident plane waves by non-circular gyrotropic cylinders. The solution method was based on the newly developed SUPER-CVWFs and the EBCM. Our suggested method was exhaustively validated for various isotropic, gyroelectric and gyromagnetic scenarios, and for various MS configurations, by comparing its results with an analytical solution and COMSOL. In addition, the superior efficiency of our approach compared to the commercial software was also demonstrated. As a potential microwave application, we studied the MS by YIG ferrite configurations, and we demonstrated broadband forward scattering by examining the effect of the oblique incidence and the MS mechanism on the scattering response, as well as the role of anisotropy in the optimization of the system. Overall, our method may be used to analyze and design contemporary microwave, optical, and photonic applications.

\appendices

\setcounter{equation}{0}
\renewcommand{\theequation}{A.\arabic{equation}}
\section{Graf's Formulas for the CVWFs}\label{APP_A}

Herein, we express Graf's formulas in vectorial notation, used for developing the MS solution. In what follows, we consider translation from ${\rm O}_qx_qy_q$ to ${\rm O}_px_py_p$. To construct Graf's formula for ${\mathbf M}^{(3)}_m(k_c,\xt_q)$, we define the generating function $\psi_m(k_c,\xt_q)=H_m(k_c\rho_q)e^{im\varphi_q}$. Then, we apply ${\mathbf M}^{(3)}_m(k_c,\xt_p)=\nabla_p\times{\mathbf e}_{zq}\psi_m(k_c,\xt_q)$, with ${\mathbf e}_{zq}\equiv{\mathbf e}_{zp}$ and $\psi(k_c,\xt_q)=\sum_{l=-\infty}^\infty e^{i(m-l)\varphi_{qp}}H_{l-m}(k_c d_{qp})J_l(k_c\rho_p)e^{il\varphi_p}$ \cite{Felbacq1994-zg}. In addition, to construct Graf's formula for ${\mathbf N}^{(3)}_m(k_c,\xt_q)$, we apply the property ${\mathbf N}^{(3)}_m(k_c,\xt_q)=1/k_0\nabla_p\times{\mathbf M}^{(3)}_m(k_c,\xt_q)$. After the necessary manipulations, we finally arrive at
\balg{A1}
\bma {\mathbf M}^{(3)}_m(k_c,\xt_q)\\{\mathbf N}^{(3)}_m(k_c,\xt_q)\ema&=(-1)^m\sum_{l=-\infty}^\infty(-1)^l e^{i(m-l)\varphi_{qp}}\notag\\
&\quad\,\times H_{m-l}(k_cd_{qp})\bma {\mathbf M}^{(1)}_l(k_c,\xt_p)\\{\mathbf N}^{(1)}_l(k_c,\xt_p)\ema\, .
\ealg
Equations~\eqref{A1} are used to translate the scattered fields ${\mathbf E}^{\rm sc}_q(\xt_q)$ from ${\rm O}_qx_qy_q$ to ${\rm O}_px_py_p$ when applying the EBCM for the lower branch of \eqref{16} and \eqref{17}.

Translations from ${\rm O}_qx_qy_q$ to ${\rm O}xy$ are needed when, for instance, the computation of the cross-sections with respect to the global ${\rm O}xy$ is required. These are obtained immediately by the following substitutions in the right-hand side of \eqref{A1}, i.e., $\varphi_{qp}\rightarrow\varphi^q$, $H_{m-l}(x)\rightarrow J_{m-l}(x)$, $d_{qp}\rightarrow\rho^q$, $J_l(x)\rightarrow H_l(x)$, $\xt_p\rightarrow\xt$, thus yielding
\balg{A2}
\bma {\mathbf M}^{(3)}_m(k_c,\xt_q)\\{\mathbf N}^{(3)}_m(k_c,\xt_q)\ema&=(-1)^m\sum_{l=-\infty}^\infty(-1)^l e^{i(m-l)\varphi^q}\notag\\
&\quad\,\times J_{m-l}(k_c\rho^q)\bma {\mathbf M}^{(3)}_l(k_c,\xt)\\{\mathbf N}^{(3)}_l(k_c,\xt)\ema\, .
\ealg

\setcounter{equation}{0}
\renewcommand{\theequation}{B.\arabic{equation}}
\section{Integrals for the System Matrix}\label{APP_B}

Equation~\eqref{20} features the integrals $I^p_{s,j\mu m}$, $s=1,2,3,4$. $I^p_{1,j\mu m}$ and $I^p_{2,j\mu m}$ are given by
\balg{}
&I^p_{1, j \mu m}=\int_0^{2\pi}\frac{e^{i(\mu-m)\varphi_p}}{\sqrt{1+\left(\rho'_p/\rho_p\right)^2}}W^p_j J_{\mu}(\chi^p_j \rho_p)\notag\\
&\times\left[\frac{im\rho'_p}{\rho_p^2}H_m(k_c\rho_p)+k_cH'_m(k_c\rho_p)\right]\rho_p{\rm d}\varphi_p\, ,\label{B1}\\
&I^p_{2, j \mu m}=\int_0^{2\pi}\frac{e^{i(\mu-m)\varphi_p}}{\sqrt{1+\left(\rho'_p/\rho_p\right)^2}}W^p_j J_{\mu}(\chi^p_j \rho_p)\notag\\
&\times\left[\frac{i\beta\rho'_pk_c}{k_0\rho_p}H'_m(k_c\rho_p)+\frac{m\beta}{k_0\rho_p}H_m(k_c\rho_p)\right]\rho_p{\rm d}\varphi_p \notag\\
&+\int_0^{2\pi}\!\!\!\frac{e^{i(\mu-m)\varphi_p}}{\sqrt{1+\left(\rho'_p/\rho_p\right)^2}}\Big\{\!\!-T^p_j\chi^p_jJ'_{\mu}(\chi^p_j\rho_p)+\frac{imS^p_j}{\rho_p}J_{\mu}(\chi^p_j\rho_p)\notag\\
&+\frac{\rho'_p}{\rho_p}\Big[S^p_j\chi^p_jJ'_{\mu}(\chi^p_j\rho_p)+\frac{imT^p_j}{\rho_p}J_{\mu}(\chi^p_j\rho_p)\Big]\Big\}\notag\\
&\times\frac{k^2_c}{k_0}H_m(k_c\rho_p)\rho_p{\rm d}\varphi_p\, ,\label{B2}
\ealg
while $I^p_{3,j\mu m}$ and $I^p_{4,j\mu m}$ are given by \eqref{B1} and \eqref{B2}, respectively, using the substitutions $W_l^p\rightarrow R_j^p$, $T_l^p\rightarrow N^p$, $S_l^p\rightarrow M_j^p$. In \eqref{B1} and \eqref{B2}, $\rho_p=\rho_p(\varphi_p)$, $\rho'_p={\rm d}\rho_p(\varphi_p)/{\rm d}\varphi_p$, with $\rho_p(\varphi_p)$ the polar equation of the boundary $\partial V_p$.

The integrals $\tilde{I}^p_{s,j\mu m}$, $s=1,2,3,4$, appearing in \eqref{21}, are given respectively by $I^p_{s,j\mu m}$, $s=1,2,3,4$, using the substitutions $H_m(x)\rightarrow J_m(x)$ and $H'_m(x)\rightarrow J'_m(x)$.

{\small

}

\end{document}